\documentclass{aastex631}
\hypersetup{linkcolor=blue,citecolor=cyan,filecolor=cyan,urlcolor=blue}
\usepackage{appendix}

\shorttitle{Investigating the Center-to-Limb Effects Using 3D Radiative Hydrodynamic Simulations}
\shortauthors{Irina~N. Kitiashvili}
\begin{document}

\title{Investigating the Center-to-Limb Effects in Helioseismic Data\\Using 3D Radiative Hydrodynamic Simulations}

\author[0000-0003-4144-2270]{Irina.~N. Kitiashvili}
\affiliation{Computational Physics Branch, NASA Advanced Supercomputing Division,\\NASA Ames Research Center
Moffett Field, CA 94035, USA}

\begin{abstract}
Full-disk observations from missions such as the SDO and SOHO have enabled comprehensive studies of solar oscillations and dynamics. Interpreting helioseismic and photospheric data is complicated by systematic center-to-limb variations. To explore the physical origin of these variations, we perform local 3D radiative hydrodynamic simulations that include effects of solar rotation to generate 24-hour synthetic time series of continuum intensity and Doppler velocity for nine viewing angles spanning from $-75^{\circ}$ to $75^{\circ}$. The simulations reveal a systematic decrease in 1D oscillation power toward the limbs and a pronounced East–West asymmetry that increases with frequency, primarily due to rotation-induced flows. Analysis of $\ell-\nu$  diagrams shows a decrease in the amplitude and width of the surface gravity ($f$) and resonant pressure ($p$) modes with increasing angular distance from the disk center. The amplitudes of the corresponding pseudo-modes with frequencies above the acoustic cut-off frequency increase in the intensity power spectra and are suppressed in the velocity spectra. The ring-diagram analysis of the simulation data further demonstrates anisotropic broadening of the modes, and the impact of the foreshortening effect on the energy distribution, and distinct differences in background noise and pseudo-mode structure between the intensity and velocity data. These results indicate that the center-to-limb effects arise from both geometric projection and physical factors such as line-formation height and potential effects of the radial differential rotation. The findings provide a framework for correcting helioseismic observations and demonstrate that realistic simulations are a powerful tool for disentangling geometric and physical biases in solar data.
\end{abstract}

\section{Introduction}

To understand and forecast global solar activity and its evolution across different scales, it is essential to monitor the dynamics of the solar interior. There are two major approaches to probing subsurface flows. The first one is based on the analysis of surface flow properties \citep[e.g.,][]{Ulrich2023} or tracking various features (e.g., magnetic elements, supergranulation). These types of studies rely on direct coupling between the surface and subsurface layers. For instance, such techniques are actively used to measure meridional flows and differential rotation at the photosphere by tracking small-scale magnetic elements or granulation \citep{Lamb2017,Roudier2018,Hathaway2022}. 
The alternative approach, known as helioseismology, suggests using acoustic waves to probe the solar interior \citep{Gough1983}. 
Previous analyses of the $\nu-\ell$ dependence of the background components of oscillations from GONG observations have shown that the background power also can provide information about the mechanism of solar oscillations \citep{Barban2004}. In particular, according to modeling and theoretical studies, the waves can be a result of the dynamics of vortex tubes and turbulent convection \citep[e.g.,][]{Unno1962,Stein1967,Kitiashvili2011,Kitiashvili2019} and propagate through the solar interior. Changes in thermodynamic properties and flows affect oscillation frequencies and acoustic travel times, allowing us to infer these changes from the surface to the bottom of the convection zone and below.
For a comprehensive overview of methods and the mathematical background, we refer the reader to the review papers by \citet{ChristensenDalsgaard2002} and \citet{Kosovichev2011}. Different helioseismic methods have been used to study various phenomena on a wide range of scales: from local, such as emergence and evolution of active regions \citep[e.g.,][]{Zhao2010,Ilonidis2011,Kosovichev2006a,Stefan2023}, to global, such as torsional oscillations \citep[e.g.,][]{Howe2020,Komm2021}. For the most recent advances in the field, we recommend the community review paper by \cite{Kosovichev2025}. 

Because our understanding of the solar interior relies on photospheric observations that require extensive data processing and assumptions, numerical modeling is essential for validating helioseismic inference methods and assessing their uncertainties and limitations. In particular, recent developments in 3D acoustic codes have enabled bridging the gap between global anelastic models, which are used to assess the capabilities of different global helioseismology techniques \citep[e.g.][]{Stejko2021,Stejko2022,Stefan2025}. The current state of the art in improving local helioseismology analysis involves using 3D realistic radiative models that reproduce the dynamics of the Sun's near-surface layers with high fidelity. For instance, this type of model has been used to investigate observational artifacts, such as the `concave Sun' effect in helioseismology \citep{Kitiashvili2015}, to validate inferences of subsurface flows from time-distance helioseismology \citep{Georgobiani2007,Zhao2007,Boening2021}, to interpret the observational center-to-limb studies \citep{Waidele2023}, to test horizontal magnetic field inferences below the photosphere \citep{Stefan2022}, and others.

The center-to-limb effects in solar observations have been studied by many authors, for example, the wavelength increase of spectral lines observed closer to the solar limb relative to the disk center and foreshortening effects \citep[e.g.,][]{Adams1910,Evershed1931,Evershed1936,Higgs1960,Howard1970,Stenflo2015,Zhao2016}. Also, it is well known that spectral line properties (e.g., line depth, width, asymmetry, bisector shape) vary with distance from the disk center, complicating the interpretation of observational data. In particular, helioseismic inversions can expose anomalies such as the `concave' or `shrinking' Sun \citep{Duvall2009,Zhao2012}, which can be reproduced through modeling of the center-to-limb variations \citep{Kitiashvili2015}. 
Investigating the impact of center-to-limb-related variations on observations and developing approaches to mitigate or eliminate them are primary objectives of many studies \citep[e.g.,][]{Zhao2016,Kashyap2025}. In particular, the center-to-limb effect introduces significant systematic errors in helioseismic inferences of the meridional circulation due to the strong frequency dependence of travel-time and phase shifts \citep{Chen2018}. Also, the ring-diagram analysis \citep{Hill1988} shows systematic errors for observations at off-central meridian longitudes and rings displacements due to rotation, as well as East-West symmetry in the gradient of the solar rotation  \citep{Greer2013, Greer2014,Jain2013,RabelloSoares2024}. Taking into account the high interest in developing forecasting techniques based on properties of the solar oscillations, e.g., to predict transient events, such as the emergence of magnetic flux with the following formation of active regions \citep[e.g.,][]{Ilonidis2011,Stefan2023,Kasapis2025}, it is essential to characterize the impact of the center-to-limb variations on helioseismic observations.

In this paper, we investigate the influence of the center-to-limb effect on oscillation power spectra in the presence of solar rotation and the Coriolis force under the quiet-Sun conditions. To perform the analysis, we run 3D radiative hydrodynamic simulations of the upper solar convection zone and the low atmosphere. Using the simulation results and a radiative transfer code, we synthesize the 6173~\AA~line observed by the Helioseismic and Magnetic Imager \citep[HMI;][]{Scherrer2012} on board the Solar Dynamics Observatory \citep{Pesnell2012} and calculate the HMI observables \citep{Couvidat2012}: the Doppler velocity and continuum intensity. These synthetic data are then used to calculate the oscillation power spectra. 

Previously, realistic 3D models were primarily used to test time-distance helioseismology \citep[e.g.,][]{Georgobiani2007,Zhao2007}. Our analysis investigates changes in the oscillation properties of the continuum intensity and Doppler velocity due to center-to-limb effects, focusing on the 3D power spectra used in the ring-diagram analysis. Using the identical simulation data to construct synthetic observables at various disk locations allows us to isolate and investigate the center-to-limb effect, which is not possible in observations. We first provide a brief description of the numerical model setup, spectral line synthesis, and Doppler velocity and intensity calculations, and discuss  general center-to-limb properties revealed in the synthetic observables (Section~\ref{sec:model}). Section~\ref{sec:l-nu_diagrams} describes properties of the surface gravity ($f$) and pressure resonant ($p$) modes as well as pseudo-modes (interference peaks above the acoustic cutoff frequency) presented in the oscillation power spectra as a function os the angular degree of spherical harmonics and frequency ($\ell-\nu$ diagrams). Then, we discuss properties of the 3D power spectra (the ring diagrams) at different longitudes (Section~\ref{sec:rings}). 
In the final Section~\ref{sec:conclusion}, we summarize our results and discuss their possible applications, as well as the impact of the foreshortening effect on the energy distribution in ring diagrams from observed Doppler velocities.

\section{Generation of Synthetic Observables and Mode Definition}\label{sec:model}

In this paper, we used 3D radiative hydrodynamic simulations of the upper convection zone and the low atmosphere, which revealed the formation of radial differential rotation, meridional flows, and a leptocline \citep{Kitiashvili2023}. The simulations were obtained using the \texttt{StellarBox} code \citep{Wray2018}, which enables the modeling of solar convection properties and dynamics with high realism. The computational domain of the model is a rectangular box 80~Mm wide and 26~Mm deep (including 1~Mm of the atmosphere). The grid spacing is 100 km in the horizontal direction and extends vertically from 25 km at the photosphere to 82 km at the bottom of the computational domain. Thus, the overall mesh size is $800\times800\times400$ (nx$\times$ny$\times$nz). The solar rotation is implemented using the f-plane approximation in the rotational frame, with a rotational period of 27~days at $30^{\mathrm{o}}$ latitude \citep{Kitiashvili2023}. 

In the analysis, we use a 24-hour-long time series of simulations with a 45-second cadence to synthesize the operational 6173\AA   ~(Fe~I) line of the Helioseismic and Magnetic Imager \citep{Scherrer2012}. The line synthesis has been performed using the \texttt{Spinor} radiative transfer code~\citep{Frutiger2000} for the reference wavelength of $\lambda_0=6173.3$~\AA, a range $\pm0.4$~\AA~with spectral resolution of $\lambda/\Delta\lambda\sim 3\times10^6$. The line synthesis was performed for nine distances from the central meridian: $\pm 75^{\mathrm{o}}$, $\pm 60^{\mathrm{o}}$, $\pm 45^{\mathrm{o}}$, $\pm 30^{\mathrm{o}}$, and $0^{\mathrm{o}}$ (disk center).
Since simulations are performed in the rotating frame, synthetic observables correspond to tracking the observed area with the imposed rotation rate. 
In this work, we use the resulting 2D time series of the continuum intensity (${\rm I_c}$) and the Doppler velocity (${\rm V_D}$). For the continuum intensity calculations, we employ the wavelength $\lambda_c=6172.9$~\AA, which is outside of the reference Fe~I line. The Doppler velocity has been computed with the center-of-gravity method
\begin{equation}
    {\rm V_D}=c*\frac{\Delta\lambda}{\lambda_0}, \quad \Delta\lambda=\frac{\sum I_i \lambda_i}{\sum I_i}-\lambda_0,
\end{equation}
where $c$ is the speed of light, and $I_i$ is the intensity at the wavelength $\lambda_i$.

\begin{figure}[b]
	\begin{center}
		\includegraphics[scale=0.9]{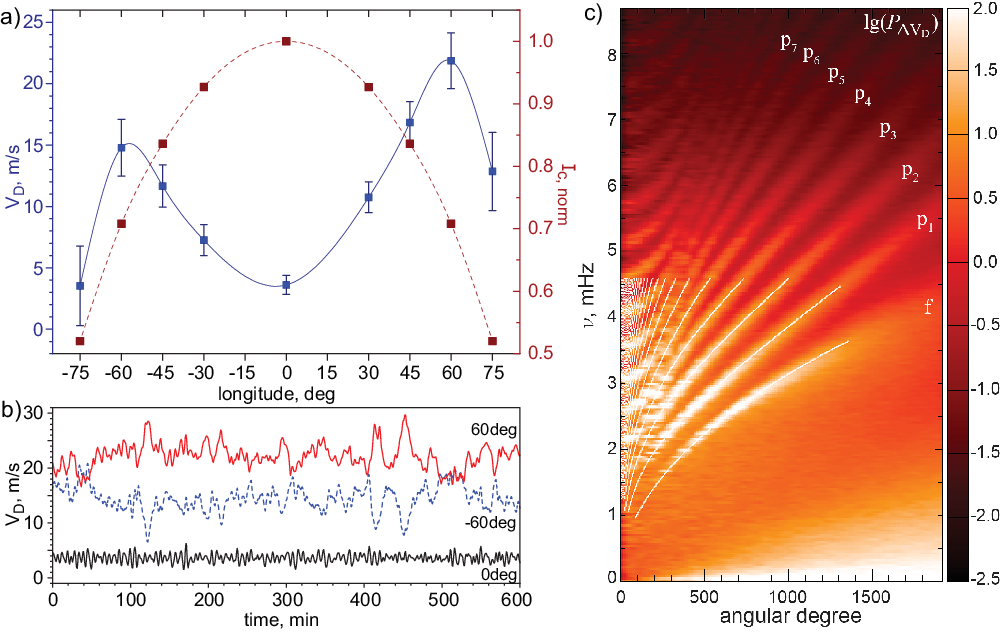}
	\end{center} 
	\caption{Panel a) Variations with longitude of the Doppler velocity (${\rm V_D}$; blue curve) and the continuum intensity (${\rm I_c}$; red curve). The vertical bars correspond to $1\sigma$ variations for the time difference of ${\rm V_D}$. Panel b: Comparison of the time variations of ${\rm V_D}$ at $\pm 60^{\mathrm{o}}$ longitudes (red solid and blue dashed curves) and the disk center (black curve) shows 5-minute variations, contribution of the solar rotation, and dynamics of the atmosphere for areas closer to the solar limb. Panel c): The logarithm of the oscillation power spectrum (the $\ell$-$\nu$ diagram) from synthetic Doppler velocity at the disk center. The  overplotted $f$- and $p$-mode frequencies (white curves)  represents results of fitting to the observed power spectra from SDO/HMI by \citep{Reiter2020}. 
    \label{fig:VdIc}}
\end{figure}

We consider two types of resonance modes: acoustic or pressure ({\it p}) modes,  which are trapped inside the Sun, and surface gravity or fundamental ($f$) modes. The modes are primarily excited by turbulent motions of the solar plasma, which excite the pressure ($p$-modes) oscillations through both entropy fluctuations and Reynolds stresses \citep{Stein1967,Goldreich1988} and localized events often associated with dynamics of vortex tubes in the intergranular lanes \citep{Kitiashvili2011}. The surface gravity modes in the observed angular degree range are confined to shallow near-surface layers. In contrast to the resonance modes, the pseudo-modes are not trapped inside the Sun. They result from the interference of high-frequency wavefronts that propagate directly from the subsurface acoustic sources and are reflected back into the atmosphere from layers deeper than the source location \citep{Kumar1990,Nigam1999}. Transition from the resonance to pseudo-modes is determined by the cut-off frequency defined as \citep{Deubner1984}
\begin{equation}
    \omega_c=\frac{c_s(r)}{2H}\sqrt{1-2\frac{{\rm d}H}{{\rm d}r}}, \quad H=\left|\frac{{\rm dlog}\rho(r)}{{\rm d}r}\right|^{-1},
\end{equation}
where $c_s$ is sound speed, $H$ is the density scale height, $\rho$ is plasma density, and $r$ is the solar radius.

\begin{figure}[b]
	\begin{center}
		\includegraphics[scale=0.62]{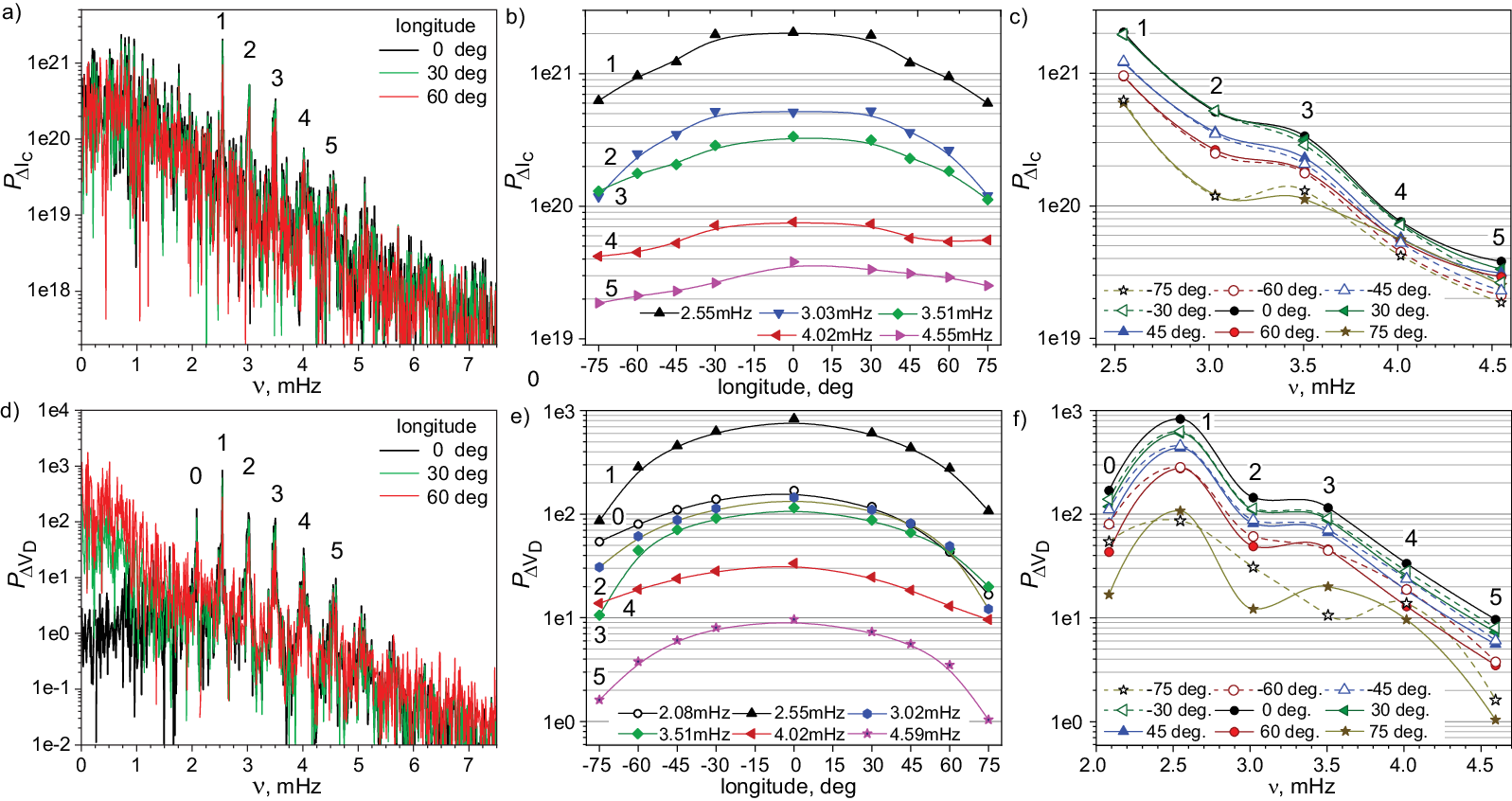}
	\end{center} 
	\caption{Center to limb variations in the power spectra (panels a and d), and change in the oscillation power of the resonance modes as a function of the longitude (panels b and e) and frequency (panels c and f) for the continuum intensity (panels a-c) and the Doppler velocity (panels d-f). The resonance modes are labeled to facilitate tracing their properties across the panels. }
    \label{fig:powIcVd}
\end{figure}

\section{General Properties of Observables at Different Longitudes}

The initial overview of the synthesized observables shows a decrease in the mean continuum intensity with increasing distance from the disk center (red squares, Figure~\ref{fig:VdIc}a), known as the limb darkening effect. The mean of the Doppler velocity increases from the disk center to $60^{\mathrm{o}}$ (blue squares, Figure~\ref{fig:VdIc}a) and shows a significant decrease for $75^{\mathrm{o}}$ longitude. Because of rotation, the Doppler velocity profile exhibits East--West asymmetry, with faster {\bf line-of-sight velocity} closer to the western limb of the Sun and a more significant velocity decrease at $-75^{\mathrm{o}}$ longitude, near the eastern limb. Since we are considering an idealized situation with dynamics identical at all longitudes and the only difference being the viewing angle within the computational domain, the mean variations in Doppler velocity are mostly anticorrelated for opposite viewing angles (Figure~\ref{fig:VdIc}b). The stronger fluctuations in the flows at greater distances from the disk center reflect a greater contribution from atmospheric dynamics.  To remove general systematic changes in the continuum intensity and Doppler velocities, we subtract the mean values from the corresponding time series for the following analysis.
Comparison of the angular degree -- frequency ($\ell-\nu$) diagram obtained for the 24-hour Doppler velocity data at the disk center and results of the ridge fitting using the full-disk SDO/HMI Dopplergrams obtained by \cite{Reiter2020} shows a very good alignment of the $f$- and the first six $p$-modes (Figure~\ref{fig:VdIc}c). Due to the limited box size, the simulations can capture modes only for $\ell > 54$. At the low $\ell$, the ridge show saturation at frequencies that diverge from observations. We analyze only the fully resolved higher-degree oscillation modes in the simulations.

Due to the changes in the properties of the continuum intensity and Doppler velocity with longitude, variations in oscillation properties are expected. In particular, for the continuum intensity (Figure~\ref{fig:powIcVd}a), the power of the spectra decreases at larger distances from the central meridian due to observing the line continuum at higher atmospheric layers \citep[e.g.,][]{Faurobert2012,Holzreuter2013,Kitiashvili2015,Waidele2023}, where temperature quickly decreases. Dependence of the oscillation power in the continuum intensity carried by each resonance mode shown in Figure~\ref{fig:powIcVd} (panels b and c) demonstrates the East-West asymmetry at higher frequencies, potentially due to the radial differential solar rotation. This trend is not valid for $\pm 75^{\mathrm{o}}$ longitudes (Figure~\ref{fig:powIcVd}c), which can be explained by a very high noise level combined with a significant foreshortening and a significant decrease of the modes' power at the higher layers of atmosphere.
Similarly to the continuum intensity, the power carried by individual modes in the Doppler velocities decreases closer to the limb. In contrast, the overall power of the spectrum increases with distance from the disk center (Figure~\ref{fig:powIcVd}d), reflecting atmospheric dynamics.

The impact of the center-to-limb effect in the presence of rotation varies for different observables. In particular, in the Doppler velocity data, the East-West asymmetry is more pronounced in the longitudinal energy distribution for the resonance mode at higher frequencies (Figure~\ref{fig:powIcVd}e) in comparison to the continuum intensity, except for the mode at 2.55~mHz, where the asymmetry is negligible. Comparison of the mode power for modeled areas at the same angular distances from the disk center shows more power for areas located closer to the East limb (panel f).
The oscillation power corresponding to $\pm 75^{\mathrm{o}}$ longitudes reveals significantly stronger relative deviations that may relate to significant contributions from atmospheric dynamics and the foreshortening effect, and also may raise a question about greater uncertainties to make a definite interpretation or conclusion.

\begin{figure}[b]
	\begin{center}
		\includegraphics[scale=0.7]{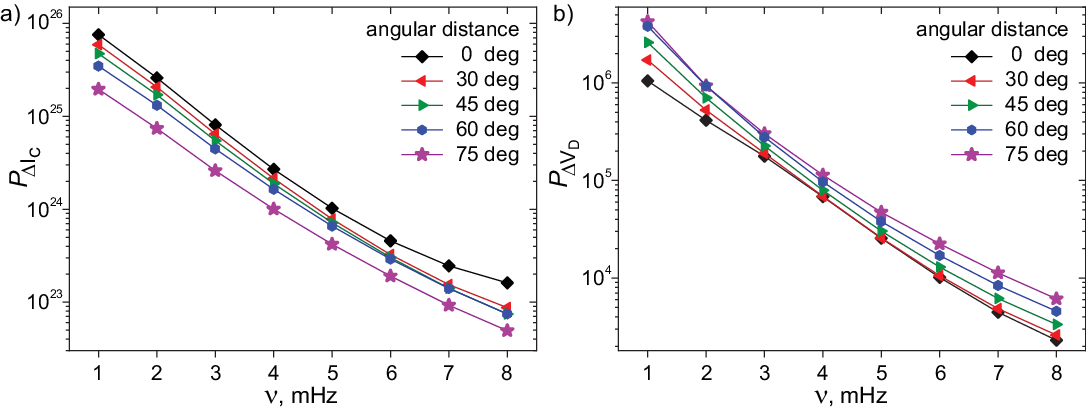}
	\end{center} 
	\caption{Variations of the acoustic power as a function of the frequency at different distances from the disk center for the continuum intensity (panel a) and the Doppler velocity (panel b) obtained by averaging the mean values of 17 power maps covering eight  1~mHz-wide frequency ranges from 0.5 to 8.5~mHz. Each power map was computed from an 8-hour time series, shifted by 1 hour. The panels show power variations only for the western hemisphere, given the similarity between the hemispheres. \label{fig:powMapsIcVd}}
\end{figure}

To quantify the oscillatory properties in more detail, we consider the power maps obtained by integrating the oscillation power in eight  sequential 1~mHz frequency ranges, from 0.5~mHz to 8.5~mHz (Figure~\ref{fig:powMapsIcVd}). For the continuum intensity, the oscillation power decreases with increasing distance from the disk center at higher frequencies. Also, above the cutoff frequencies ($\sim 5$~mHz), the power decrease rate is noticeably lower at the disk center (black diamonds, Figure~\ref{fig:powMapsIcVd}a). This effect weakens as the distance from the disk center increases. For the power maps computed from the Doppler velocity (Figure~\ref{fig:powMapsIcVd}b), the mean power decreases with the oscillation frequency and for a given frequency increases closer to the limb. At the disk center, the oscillation power decreases slightly at low frequencies (below 3.5~mHz) and more rapidly at higher frequencies. In off-center-of-disk observations, the oscillation power decreases monotonically with increasing frequency. 

\section{Longitudinal Dependence of Acoustic and Surface Gravity Modes}\label{sec:l-nu_diagrams}

\begin{figure}[b]
	\begin{center}
		\includegraphics[scale=1.35]{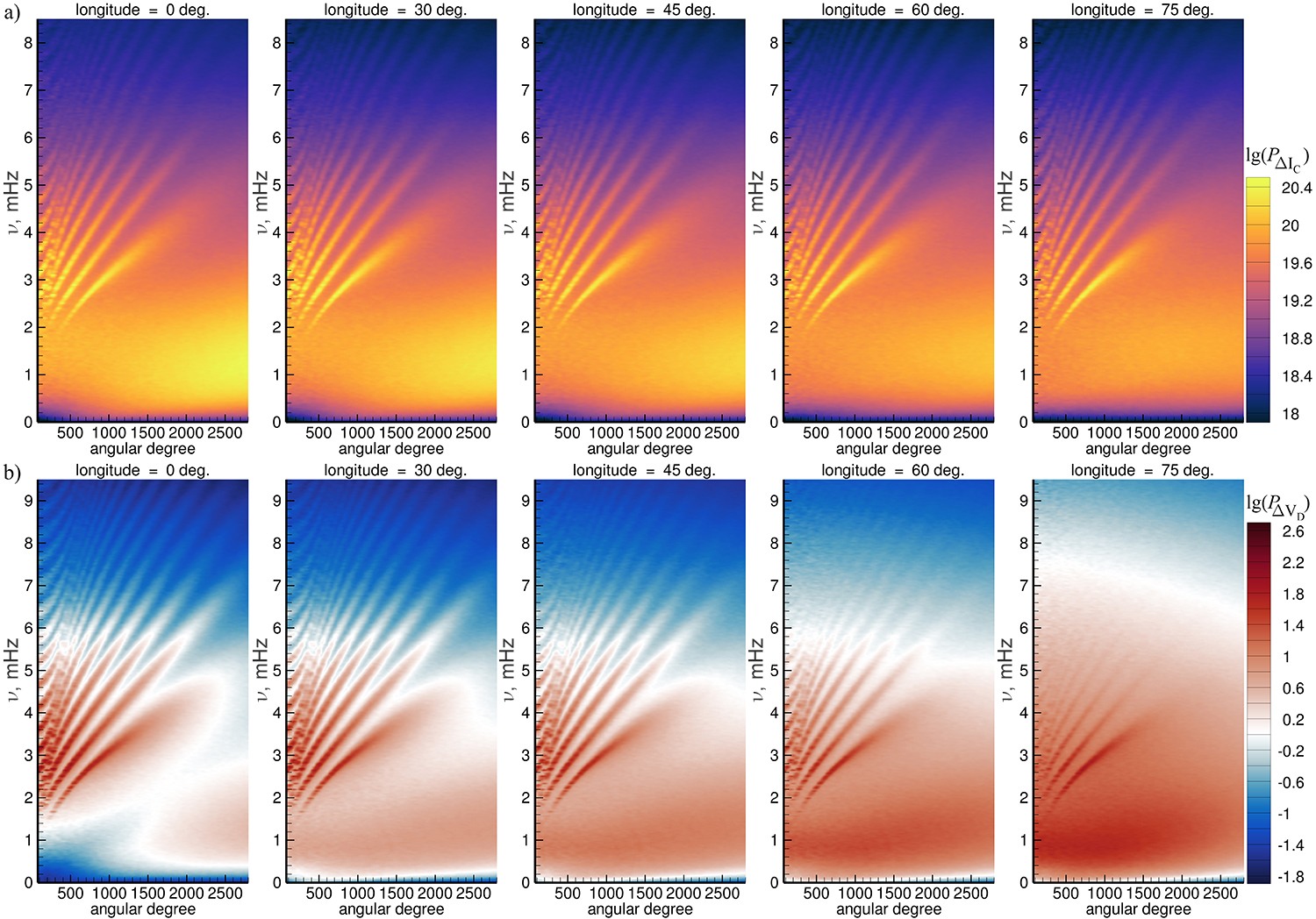}
	\end{center}
	\caption{The power spectra distribution as a function of the angular degree and frequency (the $\ell-\nu$ diagrams) obtained for the continuum intensity (panel a) and Doppler velocity fluctuations (panel b) at different distances from the disk center. The presented diagrams are the result of averaging 17 diagrams from the consecutive 8-hour time series with 1-hour time shifts. The $\ell - \nu$ diagrams are a result of integration over a constant wavenumber ($k_h=\sqrt{k_x^2+k_y^2}$) in the polar coordinates for each frequency from the 2D time series of the continuum intensity and the Doppler velocity after applying the running time differencing, and subtracting the mean values.  \label{fig:l-nu_diff}}
\end{figure}

Another way to assess the influence of the center-to-limb effect on the observed properties of solar oscillations is to analyze azimuthally averaged 3D spectra. These spectra ($\ell-\nu$ diagrams) represent the oscillation power distribution as a function of the angular degree ($\ell$) and frequency ($\nu$). The diagrams show the oscillation power carried out by turbulent convective motions (background convective noise), surface gravity ($f$) and acoustic ($p$) resonance modes, as well as the pseudo-modes  (interference peaks above the acoustic cutoff frequency), the observed properties of which vary with longitude (Figures~\ref{fig:VdIc}c,~\ref{fig:l-nu_diff}). 
To enhance the signal from resonant and pseudo-modes (appearing as ridges above the cutoff frequency) in the continuum intensity and Doppler velocity, we first subtracted the mean value from each snapshot and then calculated the temporal differences between consecutive snapshots in the time series.

Also, a 24-hour-long dataset has been partitioned into 17 subsets, each representing an 8-hour-long time series shifted by 1 hour.
Figure~\ref{fig:l-nu_diff} and the following analysis are based on the averaged diagrams. The center-to-limb variations affect the observational properties of the modes and the background noise differently for the continuum intensity (panel a) and Doppler velocity (panel b). This is related to the non-linear nature of radiative hydrodynamics in a highly stratified medium, with each observable, as well as projection and foreshortening effects, closer to the limb.

\begin{figure}[b]
	\begin{center}
		\includegraphics[scale=0.95]{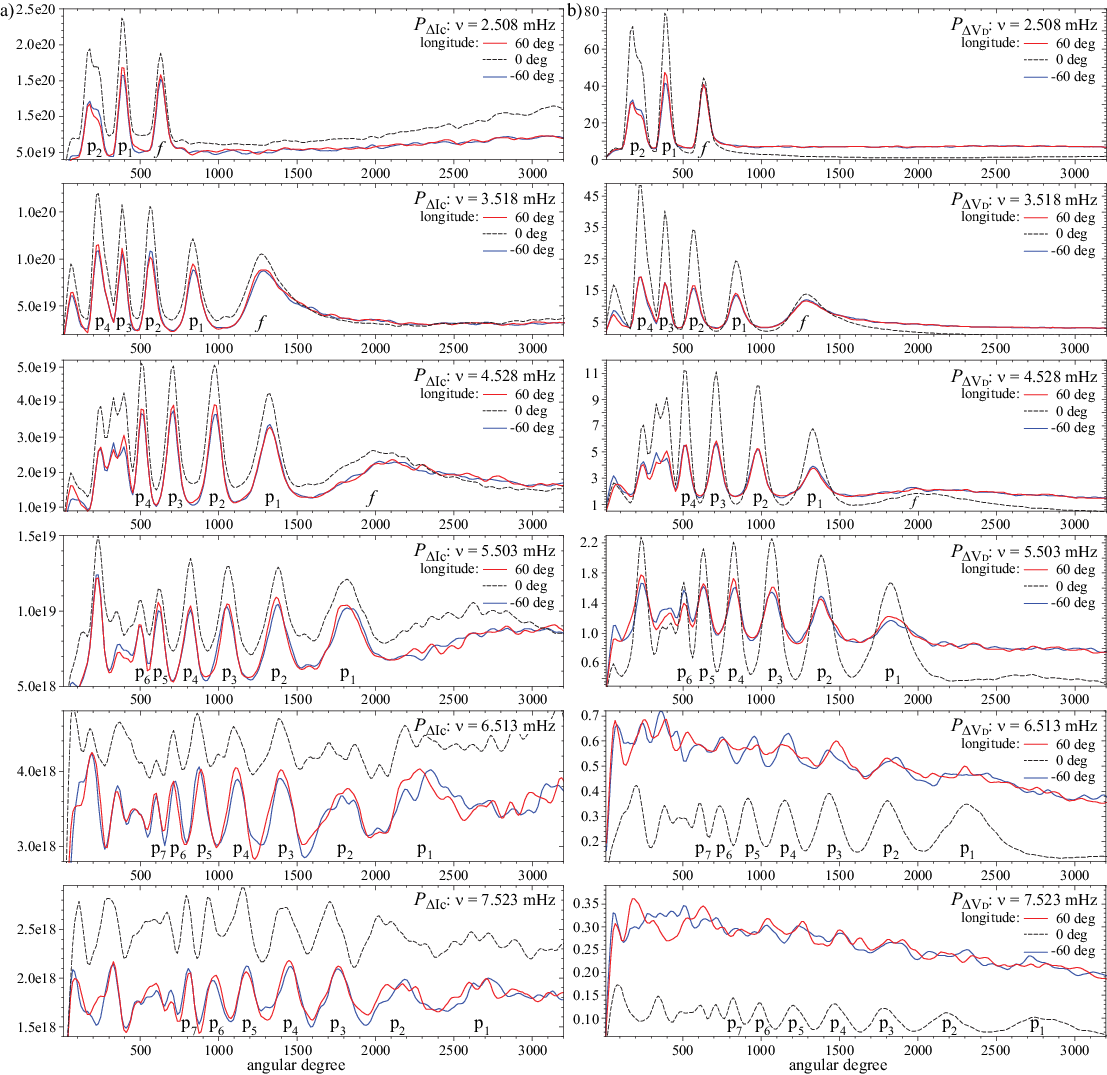}
	\end{center} 
	\caption{Comparison of the oscillation power for the continuum intensity (panel a) and Doppler velocity (panel b) at $\pm 60^{\mathrm{o}}$ longitudes toward the East (blue curves) and West (red curves) limbs at selected frequencies. The plots are extracted from the $\ell-\nu$ diagrams shown in Figure~\ref{fig:l-nu_diff}. The dashed black curves correspond to the disk center ($0^{\mathrm{o}}$) and are given as a reference. \label{fig:l-nu1}}
\end{figure}

\subsection{Continuum intensity}

The center-to-limb effect in the $\ell-\nu$ diagrams for the continuum intensity (Figure~\ref{fig:l-nu_diff}a) is manifested as a power decrease for the low-frequency convective noise, all resolved modes, and an enhancement in modes contrast obtained from the data corresponding to observations closer to the solar limb. These effects are linked to changes in the viewing angle, which cause integration through thicker atmospheric layers and increase the foreshortening, leading to the observation of the continuum intensity at higher altitudes in the solar atmosphere. Given that the amplitude of the low-frequency noise associated with photospheric temperature fluctuations decreases rapidly with height, at larger distances from the disk center, a reduction in the background noise allows tracking the oscillation modes up to $75^{\mathrm{o}}$ longitude (Figures~\ref{fig:l-nu_diff}a and~\ref{fig:l-nu1}a). This effect is particularly evident for pseudo-modes. In the $\ell$–$\nu$ diagrams constructed for off-disk-center locations, the pseudo-mode signal appears enhanced primarily due to a reduction in the background noise level, even though the absolute pseudo-mode power itself decreases. The relative comparison of modes for areas located  $60^{\mathrm{o}}$ from the disk center to the East and West limbs shows some deviations (red and blue curves in Figure~\ref{fig:l-nu1}a), which, however, are within the standard deviation (Figure~\ref{figA:sIc-Vd}a). 

\begin{figure}[b]
	\begin{center}
		\includegraphics[scale=1]{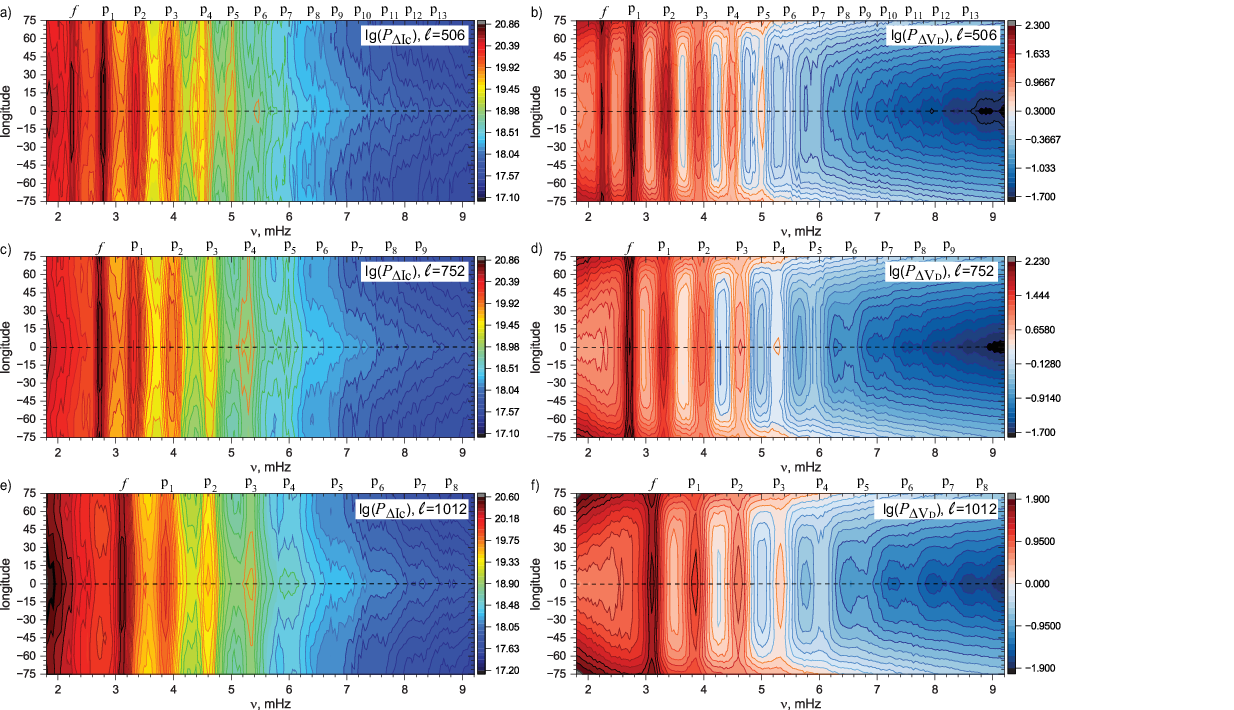}
	\end{center} 
	\caption{Frequency -- longitude diagrams of the oscillation power in the continuum intensity (panels a, c, and e) and the Doppler velocity variations (panels b, d, and f) obtained for three angular degrees $\ell$: 506 (panels a and b), 752 (panels c and d), and 1012 (panels e and f). The numbering of $p$-modes ($p_1$, $p_2$, $\ldots$) indicates the mode radial orders $n = 1, 2, \ldots$ \label{fig:l752-1012}}
\end{figure}

\begin{figure}[b]
	\begin{center}
		\includegraphics[scale=1]{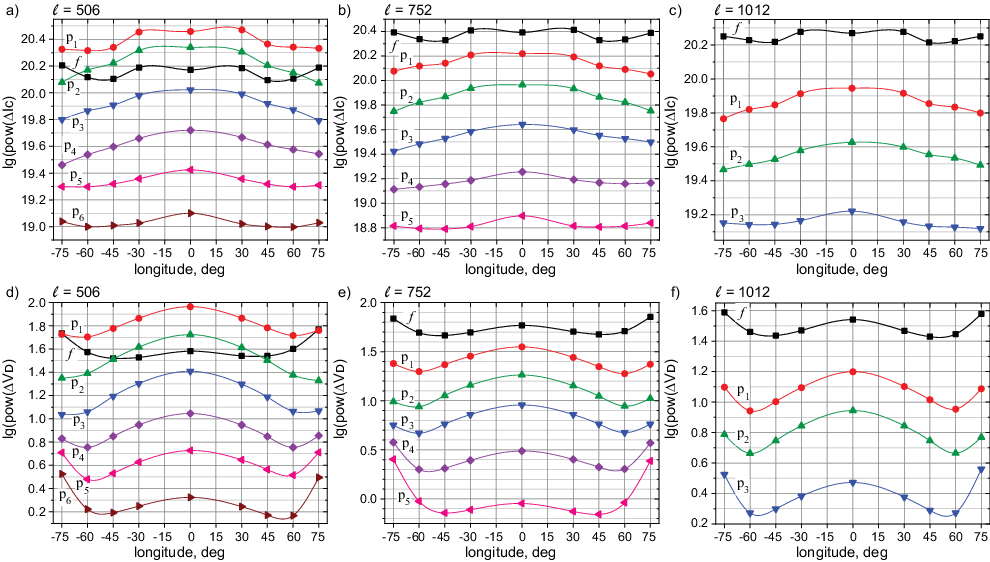}
	\end{center} 
	\caption{Power distribution with longitude for selected modes at $\ell =506$ (panels a and d), 752 (panels b and e), and 1022 (panels c and f) for the continuum intensity (panels a-c) and Doppler velocity variations (panels a-c). The modes are identified in Figure~\ref{fig:l752-1012}. \label{fig:long-l752-1012}}
\end{figure}

The structural changes in the properties of individual modes with longitude can be illustrated using the frequency-longitude diagrams for the selected spherical harmonic degree. Figure~\ref{fig:l752-1012} shows the power distribution of the continuum intensity as a function of frequency and longitude for angular degrees $\ell=$ 506, 752, and 1012. The resulting diagrams allow us to identify $f$- and several $p$-modes, revealing variations in the energy distribution among the modes at different longitudes. In contrast to the data representation in the form of classical $\ell-\nu$ diagrams (Figures~\ref{fig:l-nu_diff}a), the identification of the high-degree ridges is challenging due to the low signal-to-noise ratio and their compact location. In the frequency-longitude diagrams, the signal associated with a particular resonance mode shows the energy enhancements at the disk center (along the horizontal dashed lines in Figure~\ref{fig:l752-1012}), as discussed previously. This property allows us to identify modes up to $p_{13}$ (the radial order, $n=13$). Nevertheless, it is important to note that the retrieval of information about the properties of modes above $p_8$ is limited due to a significant weakening of the signal.

Another interesting property revealed by the frequency-longitude diagrams (Figures~\ref{fig:l752-1012}a, c, d) is a decrease in the width of the resonance modes at larger distances from the disk center. The longitudinal variation in the mode width is weaker at lower frequencies. For instance, this effect is more noticeable for high-order modes (e.g., $p_3$) than for the $f$ or $p_1$ modes. Also, the longitudinal variation of the mode width is more pronounced for higher $\ell$. The transition from the resonant $p$ modes to the pseudo-modes at about 5~mHz is manifested by a dramatic change in the energy distribution properties (starting from $p_7$ for $\ell = 506$; $p_5$ for $\ell = 752$; and $p_4$ for $\ell = 1012$; Figures~\ref{fig:l752-1012}a, c, d). At the low frequencies below $f$ mode (e.g., $\sim 2.75$~mHz for $\ell=1012$; Figure~\ref{fig:l752-1012}e), the power distribution reveals the East--West asymmetry that potentially is an effect of the radial differential rotation developed in the numerical model \citep{Kitiashvili2023}.

Analysis of the longitudinal power distribution (Figure~\ref{fig:long-l752-1012}a-c) shows its mostly symmetrical distribution relative to the disk center with a maximum at the disk center, except $p_1$ at $\ell=5-6$ (panel a). Interesting to note that the energy carried by $f$-mode at low spherical harmonic degrees (e.g., $\ell=506$, panel a) is smaller in comparison to $p_1$ and $p_2$ modes, at the disk center is weaker at the disk center and higher at $\pm 30^{\mathrm{o}}$ longitude, which probably relates to the horizontal propagation of the surface gravity waves. At present, we don't have an explanation for the asymmetry in the power distribution of the $p_1$ mode at $\ell=506$ and require further investigation.

\subsection{Doppler velocity}

The center-to-limb effect affects the properties of Doppler velocity oscillations differently from those of continuum intensity oscillations. It manifested as an increase in the background noise at larger distances from the disk center across all frequencies, leading to the weakening and disappearance of higher-degree resonant and pseudo-modes. At the same time, a signal induced by predominantly radial convective motions decreases closer to the limb due to an increase in the view angle, which causes a significant weakening of $f$ and $p$ modes (Figure~\ref{fig:l-nu_diff}b). Comparison of the power distribution at the disk center (dashed curves, Figure~\ref{fig:l-nu1}b) and $\pm 60^{\mathrm{o}}$ longitudes (red and blue curves), shows a significant decrease in the modes' power at frequencies below the cut-off frequency (2.508 -- 4.528~mHz, panel b), whereas the background noise remains relatively low. Above the cut-off frequency, the modes' amplitudes continue to decrease. However, a significant increase in background power makes them indistinguishable at high frequencies (e.g., $\nu=7.523$~mHz, Figure~\ref{fig:l-nu_diff}b). At the higher frequencies, the power of modes shows a noticeable deviation for $\pm 60^{\mathrm{o}}$ longitudes, which is smaller than the standard deviation for the corresponding modes (Figures~\ref{fig:l-nu_diff}b,~\ref{figA:sIc-Vd}b).

Similar to the continuum intensity, the distribution of the oscillation power with longitude,  computed from the Doppler velocities, shows a decrease in the mode width closer to the limb, as well as the power reduction (Figures~\ref{fig:l752-1012}b, d, f). 
In contrast to the continuum intensity, the variation of $f$-mode power with longitude is qualitatively similar to that observed for the $p$-modes (Figure~\ref{fig:long-l752-1012}d-f). The power distribution for the Doppler velocity exhibits a weak East-West asymmetry for $f$, $p_5$, and $p_6$ at $\ell=506$, which disappears at higher $\ell$.

\begin{figure}
	\begin{center}
		\includegraphics[scale=1.4]{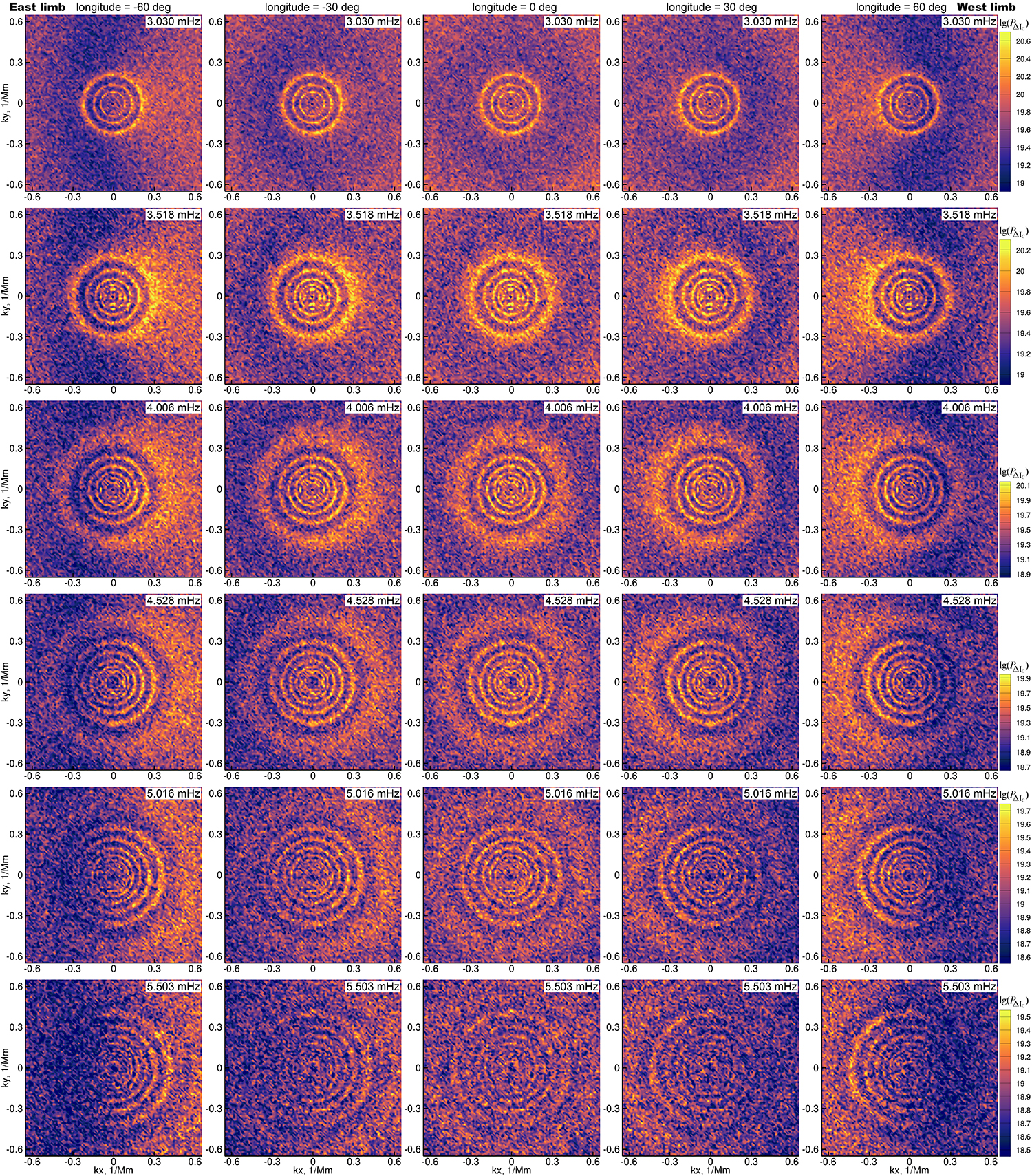}
	\end{center} 
	\caption{Ring diagrams of the continuum intensity fluctuations at five longitudes (from left to right): $-60^{\mathrm{o}}$, $-30^{\mathrm{o}}$, $0^{\mathrm{o}}$, $30^{\mathrm{o}}$, and $60^{\mathrm{o}}$ for five frequencies (from top to bottom): 3.03~mHz, 3.518~mHz, 4.006~mHz, 4.528~mHz, 5.016~mHz, and 5.503~mHz. The ring diagrams are obtained from a 24-hour data set, divided into 17 8-hour subsets using a 1-hour sliding window, and then averaged. Ring diagrams at off-disk-center locations show enhanced power toward the disk center and reduced power toward the corresponding limb.\label{fig:ringIc}}
\end{figure}

\begin{figure}[t]
	\begin{center}
		\includegraphics[scale=1.3]{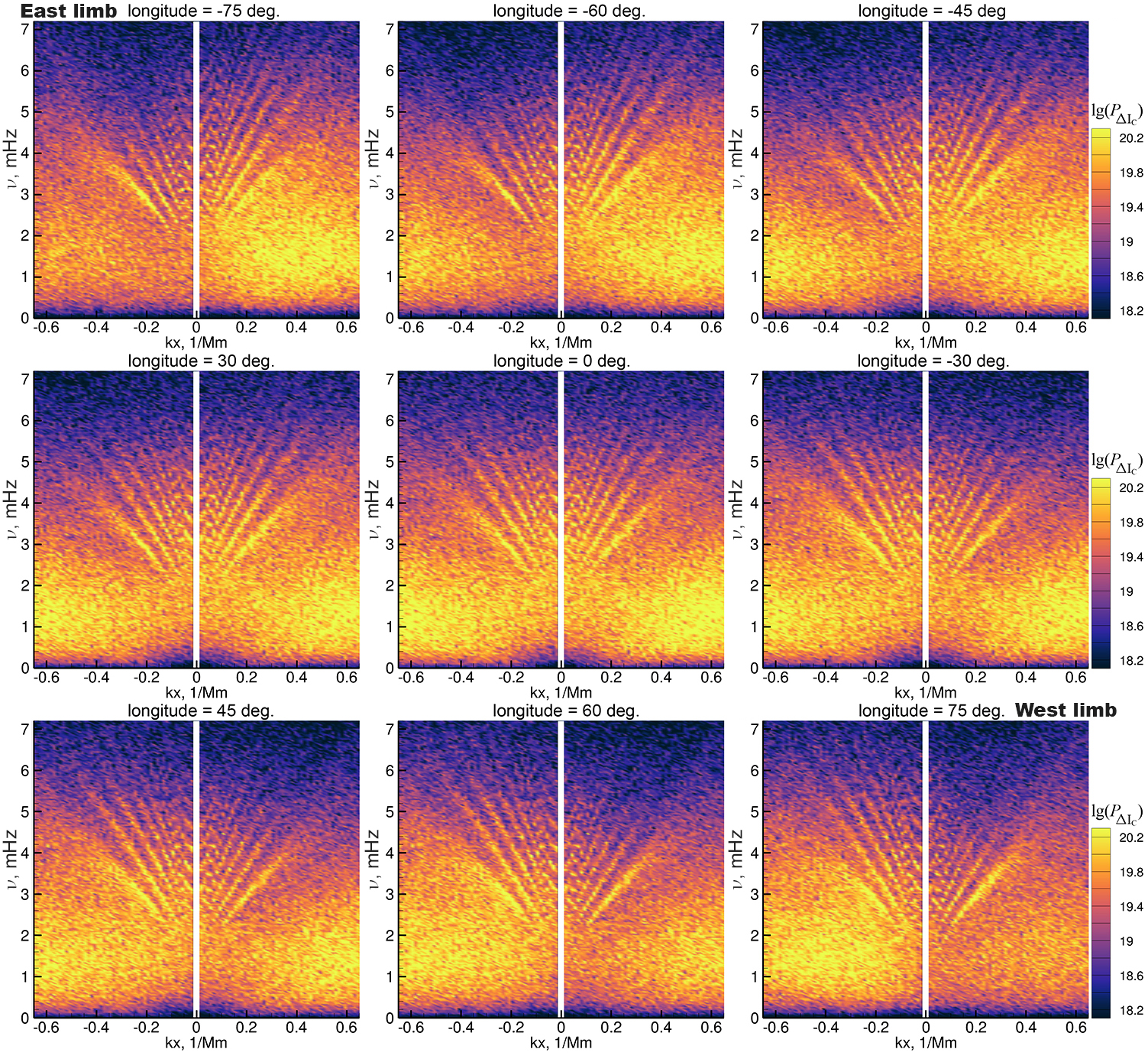}
	\end{center} 
	\caption{Distribution of the oscillations power in the continuum intensity in the azimuthal plane as a function of the wavenumber ($k_x$) and frequency ($\nu$) at nine locations on the solar disk from $-75^{\mathrm{o}}$ longitude near the East limb to $75^{\mathrm{o}}$ near the West limb.\label{fig:ring_kx-Ic}}
\end{figure}

\section{Impact of the center-to-limb effect on properties of ring-diagrams}\label{sec:rings} 

The next step in investigating the center-to-limb effect is to examine the characteristics of 3D oscillation spectra derived from the continuum intensity and Doppler velocity data. To reduce noise and remove background variations, we computed the running-time differences and subtracted the mean values from the 2D time series of synthetic observables. The resulting datasets have been split into 8-hour time series with a 1-hour sliding window to compute the 3D spectra. In this section, we consider 3D spectra averaged over the resulting 17 data cubes. To investigate the center-to-limb effect, we consider the ring diagrams that represent 2D slices in the wavevector ($k_x$, $k_y$) plane of 3D FFT power spectra ($k_x$, $k_y$, $\nu$) for a selected set of frequencies. Each slice shows an energy concentrated in the form of rings (Figure~\ref{fig:ringIc}). In each ring diagram, the outermost ring corresponds to the surface gravity ($f$) mode, and the following rings represent pressure (or acoustic, $p_1$, $p_2$, etc.) modes.  Careful modeling of ring-diagram properties is critical for inversion analyses of ring diagrams, which provide information about the structure of subsurface layers \citep[e.g.,][]{Schou1998a,Howe2015,Basu2022,RabelloSoares2024}. 

\subsection{Continuum intensity}

In this work, we are evaluating longitudinal variations of ring diagrams for six frequencies from 3~mHz to 5.5~mHz with a step of 0.5~mHz (Figure~\ref{fig:ringIc}). The center-to-limb effect in the ring diagrams obtained from continuum-intensity fluctuations shows a systematic energy redistribution driven by the oscillations and background noise. At the disk center, the energy distribution remains uniform along both the rings, which represent the surface gravity ($f$) and acoustic ($p$) modes, and in the background. However, even a relatively small shift to the East or West limbs ($\pm 30^{\mathrm{o}}$ longitude) results in the power variation within the modes, which causes an excess of power toward the disk center and a reduced power in the direction toward the corresponding limb. Note that the power variation is stronger for the outer ring (lower-order mode). These characteristics remain the same from lower to higher frequencies, which causes a significant enhancement of the signal-to-noise ratio at $\pm 60^{\mathrm{o}}$ longitudes for $p_1$, $p_2$, and $p_3$ pseudo-modes toward the corresponding limb (bottom row in Figure~\ref{fig:ringIc}). A similar variation is observed in the background noise. 
Such enhancement for the high frequency oscillations for the near limb observations (e.g., 5.5~mHz, bottom row in Figure~\ref{fig:ringIc}) can be especially important for fitting the rings close to the limb and inferring the subsurface dynamics of the Sun, whereas  utilizing the continuum intensity-based ring diagrams at the disk center can be problematic due to the high background noise. 

To investigate in more detail how the center-to-limb effect contributes to the power variations in the different directions, we consider the power spectra in the wavevector--frequency diagrams. In the azimuthal plane ($k_x-\nu$, Figure~\ref{fig:ring_kx-Ic}), the power distribution shows a systematic low-frequency suppression toward the East and West limbs and an enhancement to the disk center. 
Considering the relative power distribution among the modes, we can note that at the disk center the $f$-mode is usually weaker or comparable to the $p_1$ mode (dashed curves in Figure~\ref{figA:IccVdd}a, b). At locations closer to the limb, the $f$-mode amplitude usually remains weaker than the $p_1$ mode toward the disk center, but becomes enhanced in the limbward portion of the spectrum (Figure~\ref{figA:IccVdd}a, b). It is interesting to note that the $f$-mode power at $\pm 60^{\mathrm{o}}$ longitudes (red and blue curves in panel a) remains comparable to that at the disk center (dashed curve). At higher frequencies (e.g., 3.5 mHz; panel b), the $f$-mode power increases toward the disk center and decreases toward the limb. Moreover, the $f$-mode amplitude becomes stronger relative to the $p_1$ mode in the limb-ward direction, while remaining weaker toward the disk center.

The asymmetry in the power distribution among the modes shows a significant frequency dependence due to the center-to-limb effect. For the meridional plane ($k_y-\nu$, Figure~\ref{figA:ring_ky-Ic}), the deviations in the properties toward the equator and the North pole are not apparent and show a systematic energy decrease in the low-frequency background noise and individual ridges for the areas located closer to the East or West limb. 

\begin{figure}
	\begin{center}
		\includegraphics[scale=1.4]{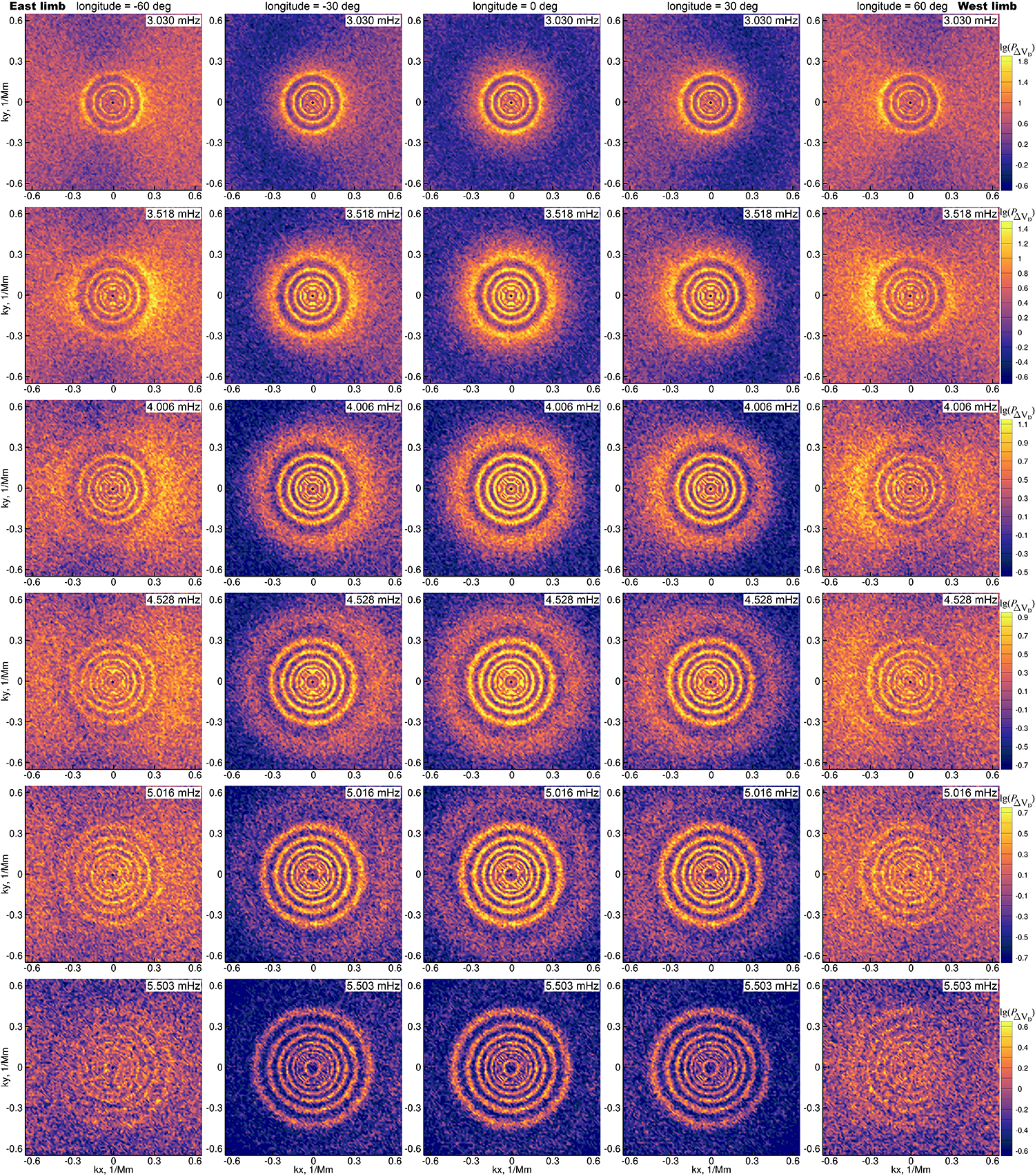}
	\end{center} 
	\caption{Ring diagrams of the Doppler velocity fluctuations at five longitudes (from left to right): $-60^{\mathrm{o}}$, $-30^{\mathrm{o}}$, $0^{\mathrm{o}}$, $30^{\mathrm{o}}$, and $60^{\mathrm{o}}$ for five frequencies (from top to bottom): 3.03~mHz, 3.518~mHz, 4.006~mHz, 4.528~mHz, 5.016~mHz, and 5.503~mHz. The ring diagrams are obtained from a 24-hour data set, divided into 17  8-hour subsets using a 1-hour sliding window, and then averaged.  Ring diagrams at off-disk-center locations show enhanced power toward the disk center.
    \label{fig:ringVd}}
\end{figure}

\begin{figure}[t]
	\begin{center}
		\includegraphics[scale=1.3]{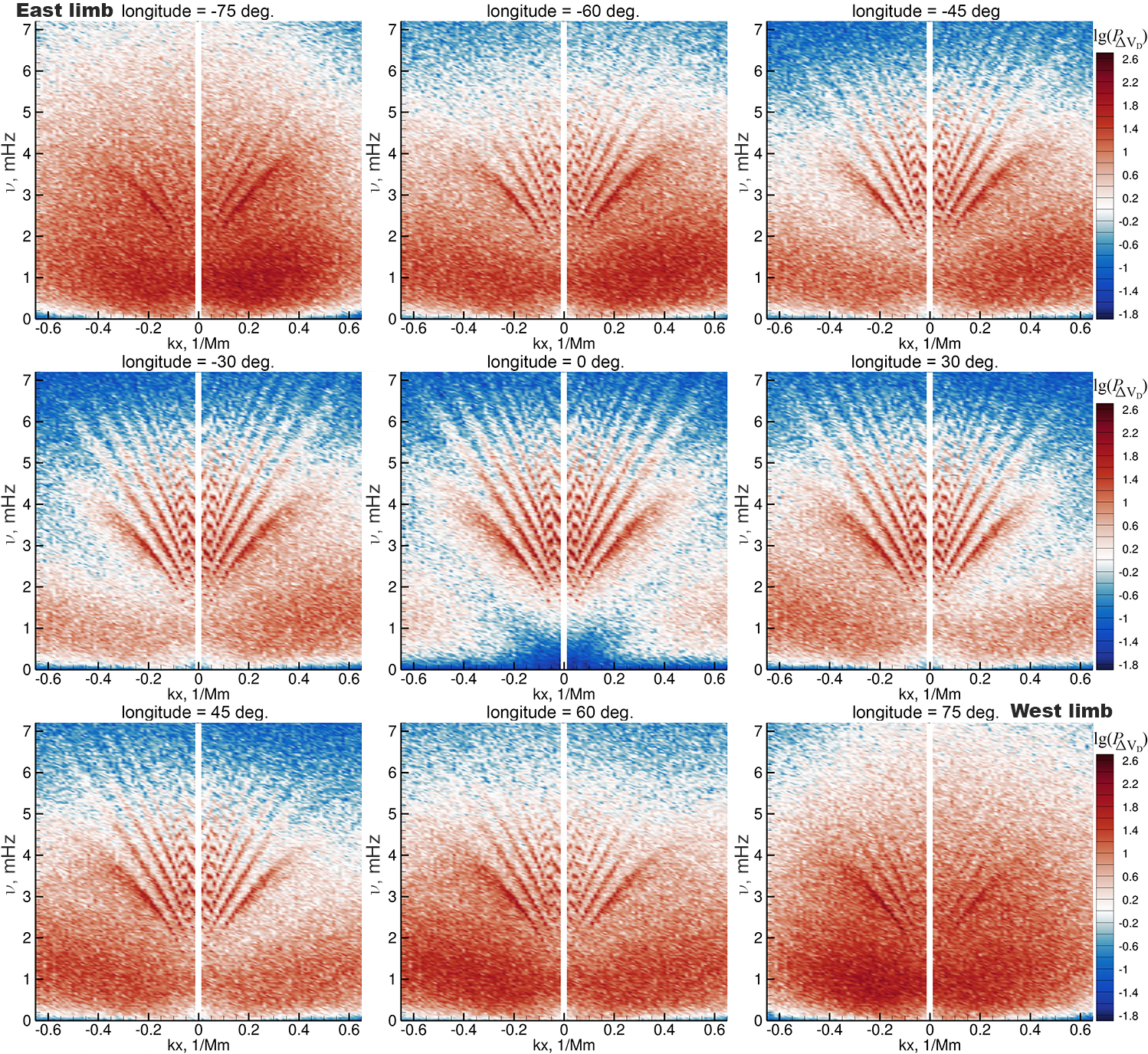}
	\end{center} 
	\caption{Distribution of the oscillation power for the Doppler velocity oscillations in the azimuthal plane as a function of the wavenumber ($k_x$) and frequency ($\nu$) at nine locations on the solar disk from $-75^{\mathrm{o}}$ longitude near the East limb to $75^{\mathrm{o}}$ near the West limb. \label{fig:ring_kx-Vd}}
\end{figure}

\subsection{Doppler velocity}

In the case of the ring diagrams computed from time series of the Doppler velocity data (Figure~\ref{fig:ringVd}), at the disk center, the resulting rings have a significantly better contrast in comparison to the continuum intensity. Similar to the continuum intensity-based ring diagrams (Figure~\ref{fig:ringIc}), the off-disk center data (e.g., $\pm30^{\mathrm{o}}$) show an enhancement of the background noise toward the disk center and the power distribution in rings (Figure~\ref{fig:ringVd}). However, because of the power reduction in the meridional plane (along $k_y$), a weaker power enhancement can be noted toward the corresponding limb. This property is most noticeable for $\nu \sim 3.5$~mHz at $\pm60^{\mathrm{o}}$ (Figure~\ref{fig:ringVd}). For the higher frequencies, $\nu > 4.1$~mHz, the mode power is suppressed toward the corresponding limb.

Similarly to the continuum intensity, we examine how the observed power of solar oscillations is affected by the center-to-limb effect by considering 3D power spectra in the azimuthal plane ($k_x-\nu$; Figure~\ref{fig:ring_kx-Vd}) and the meridional plane ($k_y-\nu$; Figure~\ref{figA:ring_ky-Vd}). In the azimuthal plane, the power distribution of the low-frequency noise becomes increasingly asymmetric, with an enhancement toward the disk center and a deficit in the opposite direction.
The surface gravity ($f$) and pressure ($p$) modes show a similar trend in the power distribution and also show the mode extension into the higher frequencies toward the disk center (Figure~\ref{fig:ring_kx-Vd}). 
The cross-comparison of the modes in the Doppler velocities (Figure~\ref{figA:IccVdd}c, d) reveals a distribution qualitatively similar to that of the continuum intensity (panels a and b), with the $f$ mode weaker than the $p_1$ mode at disk center. The power distribution at $\pm 60^{\mathrm{o}}$ longitude is also similar to that in the continuum intensity, except that the $p$-mode power is significantly reduced. In contrast to the continuum intensity, the relative power distribution between the $f$ and $p_1$ modes does not show a clear trend.

For the meridional plane ($k_y-\nu$; Figure~\ref{figA:ring_ky-Vd}), in general, properties of the background noise and ridges are mostly symmetrical. Similar to the intensity spectra, some asymmetries are evident in the power distribution of the low-frequency noise and in the ridges associated with meridional flows.

\begin{figure}[t]
	\begin{center}
		\includegraphics[scale=1.3]{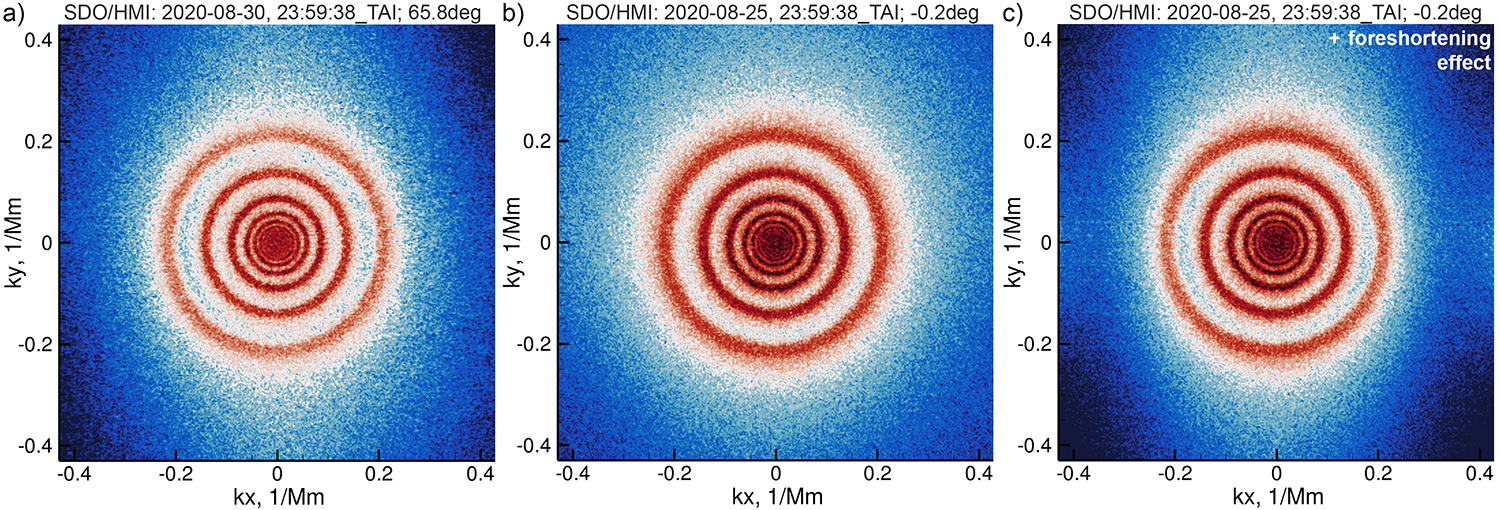}
	\end{center} 
	\caption{Ring diagrams of the Doppler velocity fluctuations at $65.8^{\mathrm{o}}$ (panel a), and  $-0.2^{\mathrm{o}}$ (panels b), and $-0.2^{\mathrm{o}}$ with artificially added foreshortening effect (panel c). The ring diagrams are obtained from an 8-hour time series of $30.6^{\mathrm{o}}\times30.6^{\mathrm{o}}$ areas of SDO/HMI observations, tracked with the Carrington rotation rate at $30^{\mathrm{o}}$ latitude. To reduce noise, the ring diagrams are averaged over a 0.1~mHz frequency range centered at 3~mHz. Because of the ring diagram's low power at large distances from the disk center, the color range in panel a) is adjusted to highlight the power enhancement along the meridional plane ($k_y$). The color range on panels b) and c) is identical.} \label{fig:ring_hmi}
\end{figure}

\section{Discussion and Conclusions}\label{sec:conclusion}

The center-to-limb effect is known as a major source of systematic variations in the helioseismic inferences \citep[e.g.,][]{Duvall2009,Zhao2012}. Previously, it was demonstrated that the observed variations in the observed signal are complex coupling effects of radiation, turbulent flows at different atmosphere layers \citep[e.g.,][]{Kitiashvili2015} manifested as power variations, convective blueshift, variations of helioseismic travel-times, as well as the ``concave” Sun effect. Therefore, understanding the physical nature and manifestations of the center-to-limb effect allows us to go beyond a simplistic removal of residuals \citep{Zhao2012} and to enable physics-informed improvements in the accuracy and interpretability of helioseismic inferences closer to the limb, as well as to reveal details of subsurface dynamics at high latitudes and polar regions. One possible explanation for the observed systematic variations in the helioseismic inferences is a change in the relative distribution of narrow downflows, associated with intergranular lanes, and upflows \citep{Baldner2012} due to variations in the viewing angle. The following studies have shown that photospheric observations near the solar limb correspond to higher atmospheric layers \citep[e.g.,][]{Faurobert2009,Faurobert2012,Kitiashvili2015}.

To better understand the impact of the center-to-limb effect, we employed 3D radiative hydrodynamic simulations \citep{Kitiashvili2023} to generate 24-hour time series of continuum intensity and Doppler velocity at 9 longitudes spanning $-75^{\mathrm{o}}$ to $75^{\mathrm{o}}$. This setup allows a more direct interpretation of results and links them to observables formed under identical conditions but viewed from different angles, effectively isolating geometric and radiative contributions from intrinsic dynamical effects, and provides a controlled environment to investigate how viewing angle influences the measured solar oscillations and background noise.

The analysis reveals systematic longitudinal variations in both continuum intensity and Doppler velocity power spectra, leading to a pronounced dependence of the oscillation power on the viewing angle. The overall oscillation power decreases toward the limbs, with a clear asymmetry between the eastern and western hemispheres that becomes stronger at higher frequencies (Figure~\ref{fig:powIcVd}). The integrated oscillation power over a 1~mHz band shows a weakly nonlinear dependence on longitude, with a stronger center-to-limb effect above the acoustic cut-off frequency (Figure~\ref{fig:powMapsIcVd}).

Comparison of the $\ell$–$\nu$ diagrams derived from both observables shows a consistent reduction in the power of $f$ and $p$ modes and the low-frequency noise with increasing distance from the disk center (Figure~\ref{fig:l-nu_diff}). However, a different behavior is found for the pseudo-modes: in the continuum intensity, these modes become more pronounced toward the limb, whereas in the Doppler velocity, they are gradually suppressed and merge with the background noise. Analysis of the power distribution with longitude as a function of frequency for selected angular degrees ($\nu-$longitude diagrams, Figures~\ref{fig:l752-1012} and~\ref{fig:long-l752-1012}) reveals that the resonance-mode width decreases with increasing distance from the disk center. It is interesting to note that for $\ell=506$, the $f$ mode is weaker than $p_1$ and $p_2$ modes for both continuum intensity and Doppler velocity (Figure~\ref{fig:long-l752-1012}). Also, the energy distribution along longitude for the continuum-intensity ring diagrams shows higher energy at $\pm 30^{\mathrm{o}}$ longitude than at the disk center, whereas the Doppler velocity distribution shows a gradual power decrease from the disk center close to the limb. The power increase at higher longitudes ($\pm 60^{\mathrm{o}}$) potentially indicates the significance of the foreshortening effect. This qualitative deviation indicates a different response of the velocity and continuum intensity to the propagation of surface gravity waves and warrants further investigation, including their excitation mechanism.

The ring-diagram spectra further highlight the center-to-limb asymmetry in both observables. The mode power and the background noise are systematically enhanced toward the disk center and reduced toward the corresponding limb (Figures~\ref{fig:ringIc},~\ref{fig:ringVd}). Despite their overall similarity, three significant differences between the Doppler velocity and continuum-intensity ring-diagram data are evident. First, the intensity spectra are noisier at the disk center, especially at high frequencies ($\nu \gtrsim 5.5$~mHz), where the rings become indistinct, while the Doppler-velocity spectra still show clearly defined features. Second, at larger angular distances ($\pm 60^{\mathrm{o}}$), the continuum intensity ring-diagrams exhibit asymmetric extensions of the pseudo-mode power toward the disk center (Figure~\ref{fig:ring_kx-Ic}). The enhancement of the pseudo-modes in the continuum intensity spectra allows for the fitting of rings for corresponding modes close to limb, e.g., at $\pm 60^{\mathrm{o}}$ longitude. Notably, the increase in continuum intensity power toward the limb was predicted theoretically by \citep{Toutain1999} and later confirmed by \citep{Toner1999} using SOHO/MDI limb observations, which demonstrated a significant enhancement in the mode power along with a corresponding decrease in the background level. 

In contrast, for the Doppler velocity, the pseudo-modes are best identified at the disk center. In the case of off-disk-center observations, power variations are qualitatively similar to continuum intensity, but increased noise makes it challenging to identify the modes in the areas close to the limb (Figure~\ref{fig:ringVd}). This suggests a potential improvement to the ring-diagram analysis for observations obtained closer to the limb by developing an analysis procedure that combines Doppler velocity and continuum intensity observations. Finally, the Doppler velocity-based ring diagrams show some power enhancement for $f$ modes toward the corresponding limb due to the power decrease in the equator-northern pole direction (along $k_y$) for frequencies below 4~mHz, in addition to the primary power enhancement toward the disk center (Figure~\ref{fig:ringVd}).
Interesting to note that at low frequencies (e.g., 2.5 mHz; Figure~\ref{figA:IccVdd}a), the $f$-mode power exhibits a weak center-to-limb dependence in the continuum intensity. At the disk center, the $f$-mode power is generally lower than that of the $p_1$ mode. However, at $60^{\mathrm{o}}$ from disk center, the $f$-mode power becomes higher than the $p_1$-mode power in the limb-ward part of the spectrum, while remaining lower in the disk-center-ward direction (Figure~\ref{figA:IccVdd}a,b). Similar to the continuum intensity, the power distribution as a function of $k_x$ and $\nu$, shows the power enhancement in the turbulent motions toward the disk center (Figure~\ref{fig:ring_kx-Vd}). However, because of a significant increase in noise, the pseudo-modes cannot be resolved at $\pm 60^{\mathrm{o}}$ from the disk center. Interesting to note that the $f$ mode in off-disk-center data is stronger and wider toward the disk center, whereas in the opposite direction the mode remains easily trackable despite a smaller amplitude (Figure~\ref{fig:ring_kx-Vd},~\ref{figA:IccVdd}d).
Considering the power distribution for the continuum intensity and the Doppler velocity in the meridional plane ($k_y-\nu$, Figures~\ref{figA:ring_ky-Ic},~\ref{figA:ring_ky-Vd}), a weak asymmetry between the equatorward and Northern pole directions can be noted, which may be related to meridional flows \citep{Kitiashvili2023} and requires a detailed analysis.

Comparison of simulation results with observations shows somewhat similar inhomogeneity in the distribution of power in the ring diagrams obtained from the Doppler velocity GONG observations \citep[see Figure 3 in ][]{Jain2013}, where the rings are getting weaker and partially unresolved at higher distances from the disk center, which makes it challenging to use them for inferring subsurface flows. The primary contradiction between the presented results and GONG observations lies in the orientation of the energy suppression. The results from the presented numerical model show a power enhancement along the disk-center -- limb line, whereas observations show power suppression. Such a contradiction can be attributed to differences in the examined spectral line (GONG uses the Ni line, which forms higher in the atmosphere), in spatial resolution, and in the specifics of GONG instrumentation. This is supported by later ring-diagram analysis based on previous SDO/HMI observations, which shows the power enhancement in the direction from the disk center to limb \citep[see Figure 1c in][]{Greer2014}. 
To review the issue, we used 8-hour SDO/HMI Doppler velocity maps covering a $30.6^{\mathrm{o}}\times 30.6^{\mathrm{o}}$ area, tracked at the Carrington rotation rate with the central part of the area corresponding to $65.8^{\mathrm{o}}$ longitude and the disk center. In general, the SDO/HMI-based ring diagram for data near the solar limb is in agreement with previous studies (e.g., GONG) and exhibits increased thickness and power in the equator -- northern pole direction (Figure~\ref{fig:ring_hmi}a) that is largely affected by the foreshortening effect. This asymmetry in the power distribution is absent in the ring diagram at the disk center (panel b). To artificially introduce the foreshortening effect into the data at the disk center, we degraded the spatial resolution in the azimuthal direction to mimic the coarser resolution near the limb, then rebinned the data back to the original grid. This manipulation caused partial loss of information from the original data in the azimuthal direction, resulting in a power redistribution along the rotation axis (Figure~\ref{fig:ring_hmi}a). Thus, this simple experiment reveals the impact of the foreshortening effect on the resulting power distribution in observations. 
In the synthetic observables, the foreshortening effect is only partially accounted for. On the one hand, the 1D atmospheric models in the \texttt{Spinor} code are projected onto the user-specified viewing angle, which causes projection-related stretching of photospheric structures, preventing information loss due to projection. On the other hand, in the actual observations near the limb, an analyzed area is larger in the azimuthal direction, whereas in the model, it remains the same. This suggests a potential direction for future modeling of foreshortening to develop a methodology for more accurate processing of near-limb observations.

The {\ bf results presented in the paper} demonstrate that the center-to-limb effect is not purely geometric and also depends strongly on the line-formation height and the sensitivity of the observable. The Doppler signal, which traces plasma motion, and the continuum intensity, which reflects radiative transfer effects, respond differently to viewing geometry and local flows. The observed asymmetries in mode power, width, and pseudo-mode distribution provide valuable diagnostics for correcting full-disk observations from instruments such as SDO/HMI and Solar Orbiter/PHI, as well as provide an important insight into the more complex coupling of the center-to-limb effect and observed oscillations that require taking into account in improving data analysis techniques and interpretation results. Moreover, this study highlights the potential of realistic 3D radiative hydrodynamic simulations as a powerful framework for disentangling the physical and geometric contributions to systematic observational biases. Future extensions of this work will include the effects of magnetic fields and multiple spectral lines to quantify how magnetic activity and line-formation height influence the center-to-limb behavior of solar oscillations across different latitudes, as well as a more detailed analysis of the radial differential rotation and meridional flows already present in models  and comparison with available SDO/HMI, Hinode/SOT, and Solar Orbiter/PHI observations.

\section*{Acknowledgments} I gratefully acknowledge Prof. Alexander Kosovichev for providing SDO/HMI data tracked at the Carrington rotation rate and for valuable discussions. Modeling and data analysis of the resulting data have been performed using the NASA Ames Supercomputing Facility. The presented investigation has been supported by the NASA Heliophysics Guest Investigator - Open Program (23-HGIO23\_2-0077), and the Science DRIVE (Diversify, Realize, Integrate, Venture, Educate) Center Program (COFFIES Project ``Consequences of Fields and Flows in the Interior and Exterior of the Sun''; 80NSSC22M0162).  

\newpage
\section*{Appendix A}
\renewcommand{\thefigure}{A\arabic{figure}}
\setcounter{figure}{0}
\begin{figure}[h]
    \centering
    \includegraphics[scale=1.5]{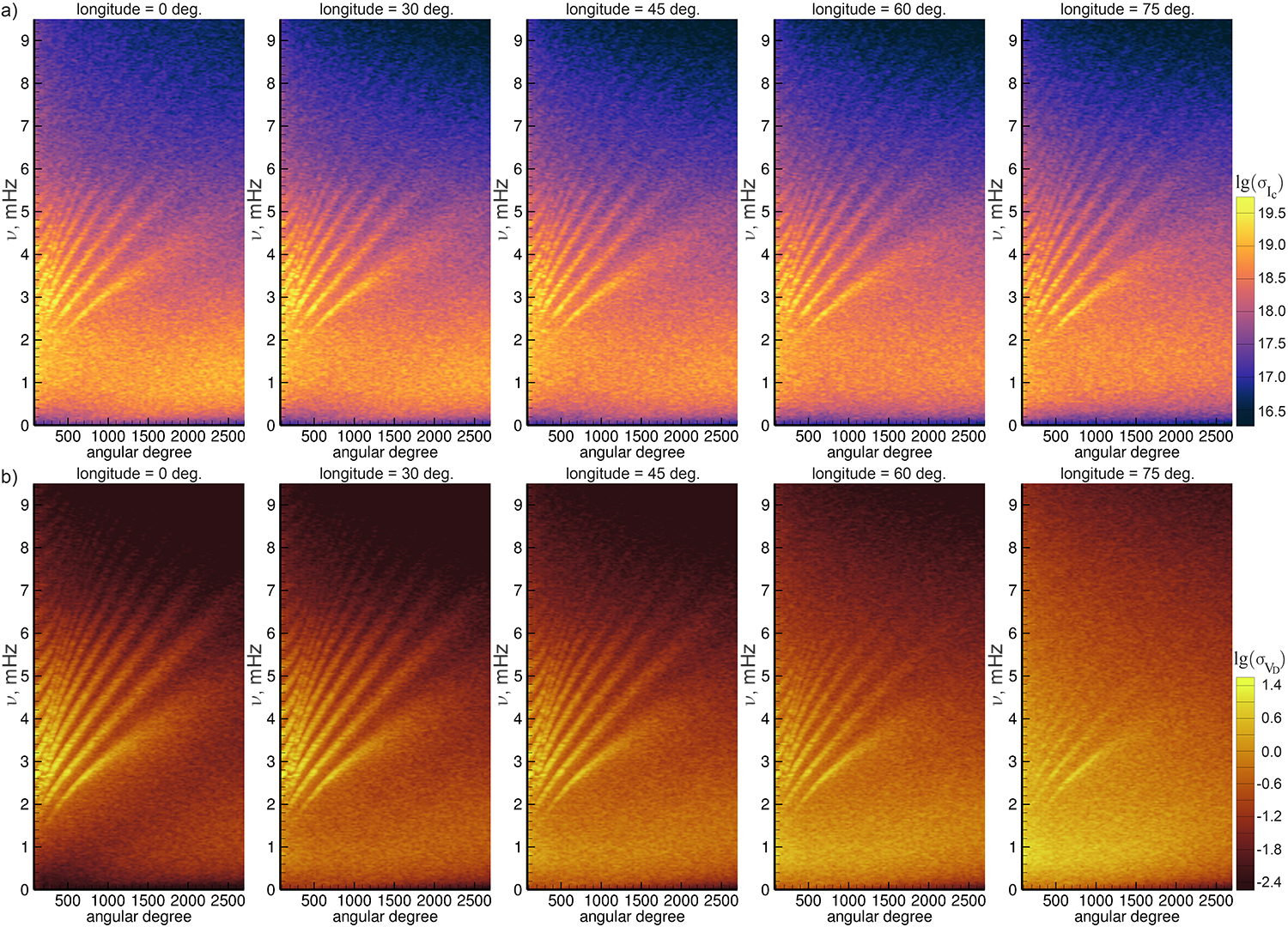}
    \caption{Distribution of the standard deviation as a function of the angular degree and frequency for the power in the continuum intensity (panel a) and the Doppler velocity (panel b) at different longitudes from the disk center (left-most panel) to $75^{\mathrm{o}}$ longitude near the West limb.}\label{figA:sIc-Vd}
\end{figure}

\begin{figure}[h]
	\begin{center}
		\includegraphics[scale=1.3]{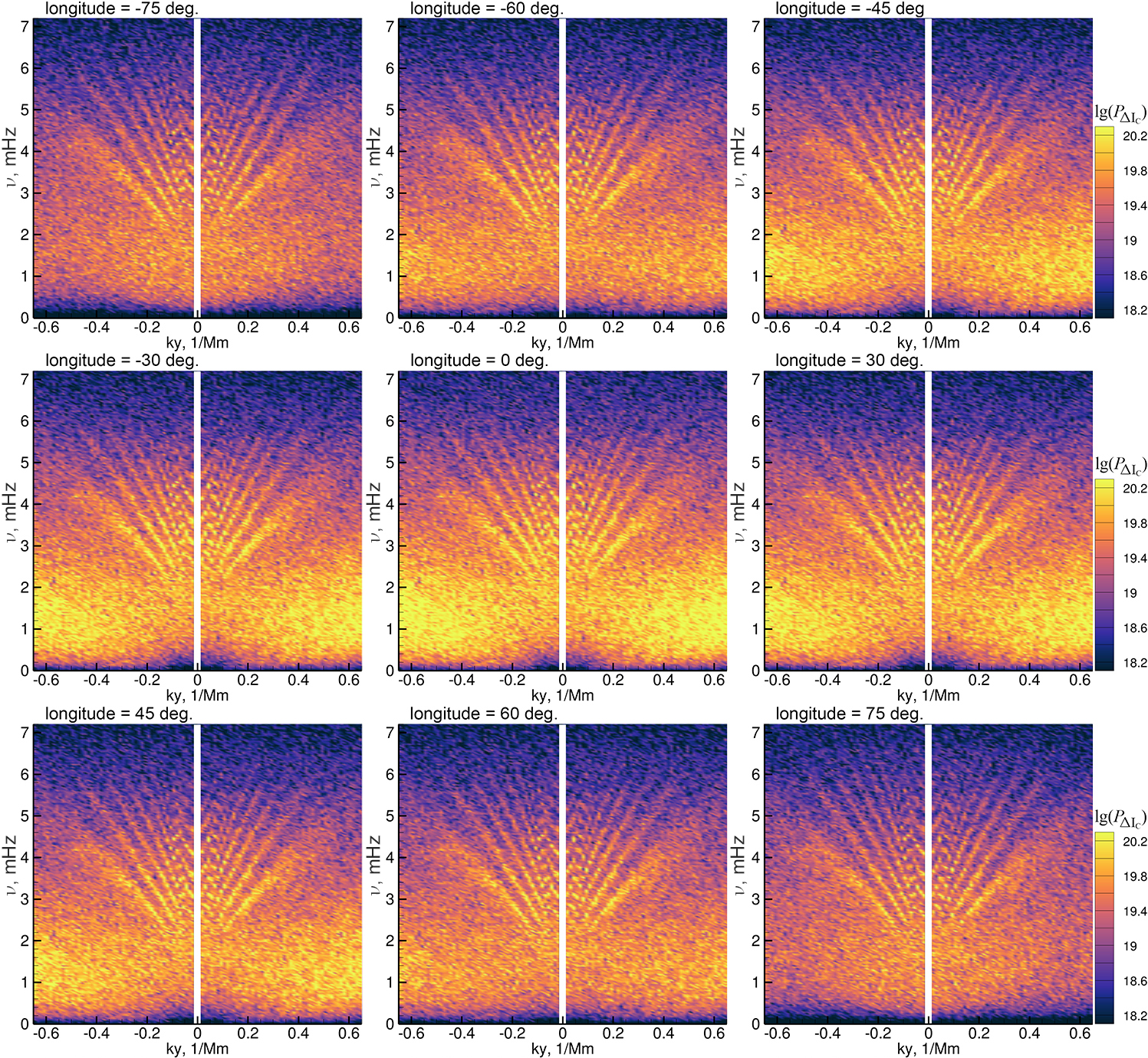}
	\end{center} 
	\caption{Distribution of oscillations in the continuum intensity in the meridional plane as a function of the wavenumber ($k_y$) and frequency ($\nu$) at nine locations on the solar disk from $-75^{\mathrm{o}}$ longitude near the East limb to $75^{\mathrm{o}}$ near the West limb. \label{figA:ring_ky-Ic}}
\end{figure}

\begin{figure}[h]
	\begin{center}
		\includegraphics[scale=1.]{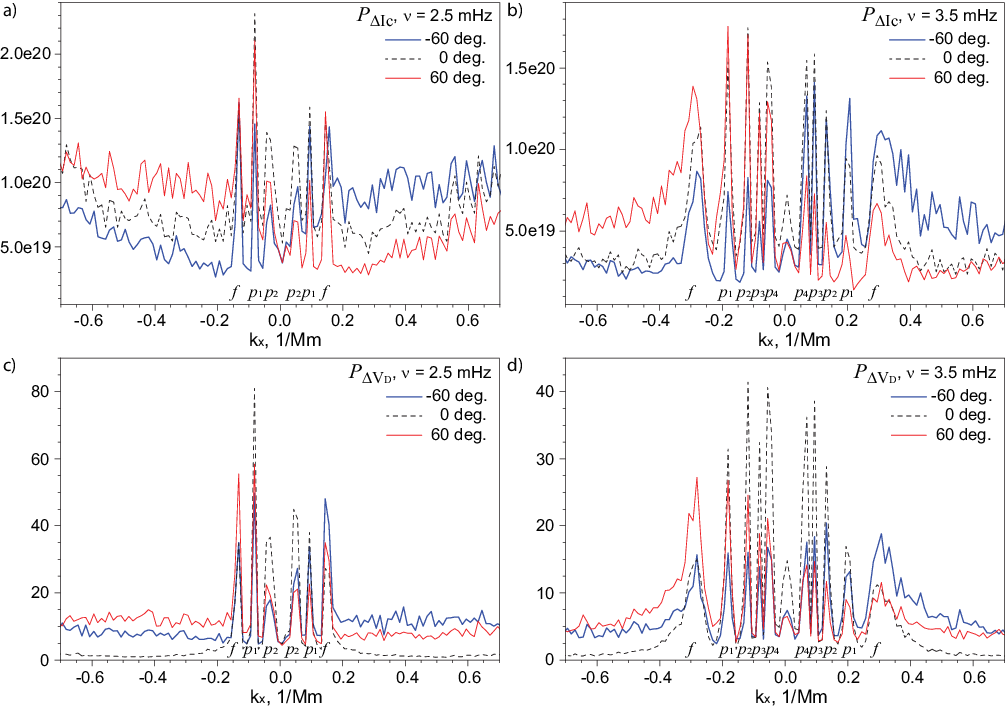}
	\end{center} 
	\caption{The power distribution in the intensity continuum (panels a, b) and Doppler velocity (panels c, d) fluctuations in the azimuthal plane for 2.5~mHz (left panels) and 3.5 mHz (right) for three longitudes: $-60^{\mathrm{o}}$ (blue curves, near East limb), $0^{\mathrm{o}}$ (black dashed curves, disk center), and $60^{\mathrm{o}}$ (red curves, near West limb). The power spectra have been extracted from the 3D power spectra shown in  Figures~\ref{fig:ringIc} and~\ref{fig:ringVd}. To reduce noise, each power spectrum was averaged over 0.1~mHz around the indicated frequency, and over the wavenumber $\Delta k_y=0.05$~1/Mm. \label{figA:IccVdd}}
\end{figure}

\begin{figure}[h]
	\begin{center}
		\includegraphics[scale=1.3]{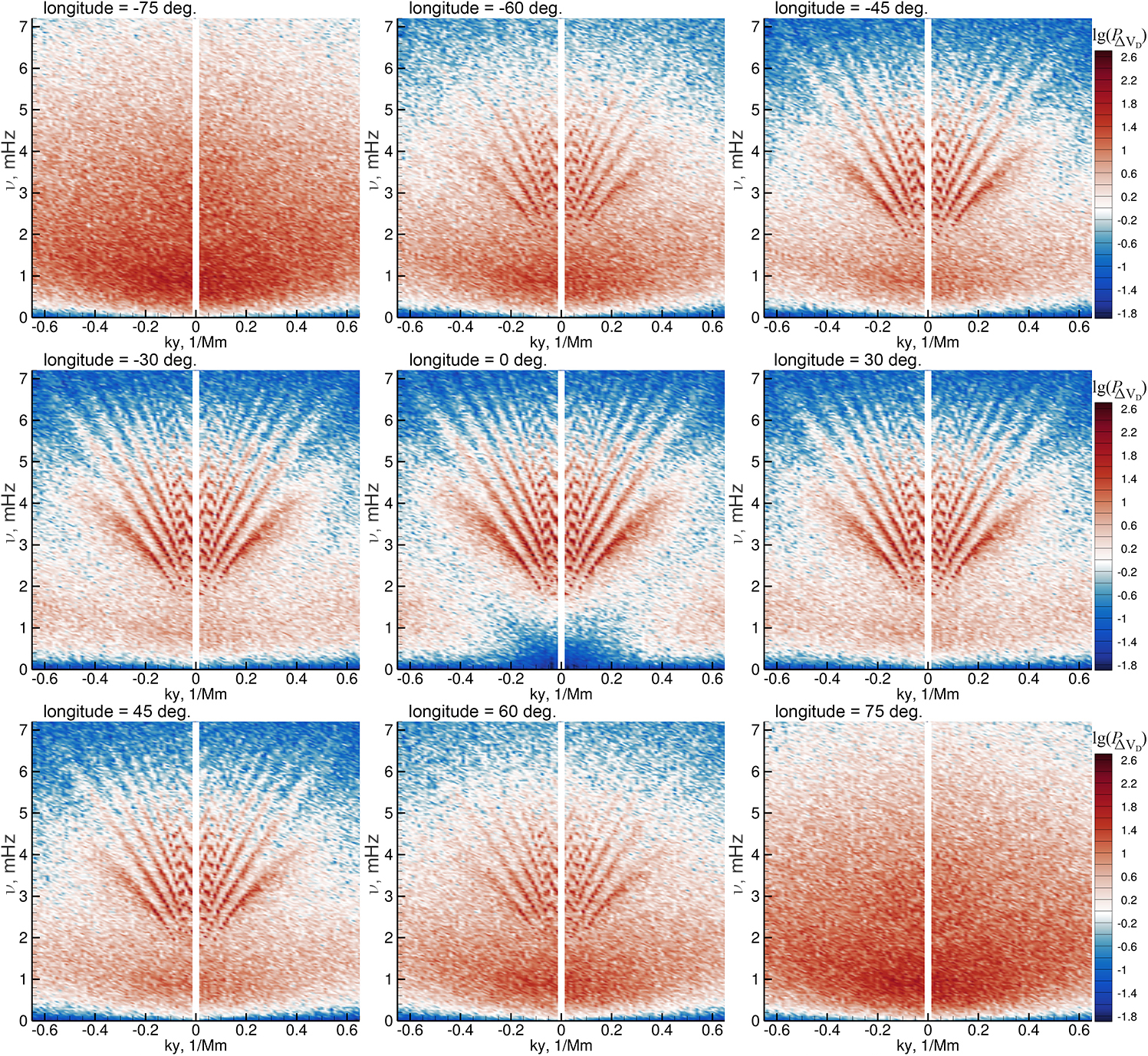}
	\end{center} 
	\caption{Distribution of oscillation power in the Doppler velocity in the meridional plane as a function of the wavenumber ($k_y$) and frequency ($\nu$) at nine locations on the solar disk from $-75^{\mathrm{o}}$ longitude near the East limb to $75^{\mathrm{o}}$ near the West limb. \label{figA:ring_ky-Vd}}
\end{figure}
\end{document}